# A Physical Interpretation of Milky Way Galaxy Dynamics from Precision Astrometrics

Jeffrey M. La Fortune
1081 N. Lake St. Neenah, WI 54956 forch2@gmail.com
25 August 2020

*Abstract*
The dynamical and virial mass of the Milky Way galaxy is estimated using latest high precision stellar halo and dwarf galaxy satellite kinematics. The new data suggest the Galaxy is a highly compact, classically thermalized object. Kinematics exhibit significant velocity-spatial substructure, distinctive dynamic partitions, and strong Keplerian signatures that run counter to popular notions of featureless and massively extended dark matter halos. The effective local escape velocity profile of the Galaxy is quantified in terms of distribution and kinematics to reveal the physics responsible for the Mass Discrepancy-Acceleration (MDAR) and Radial Acceleration (RAR) relations.

*Introduction*
With precision kinematic observations (SSDS and Gaia) made recently available, there has an interest in using this data to reduce historical uncertainty in Milky Way ΛCDM-based dark halo mass estimates. With this improvement, we present a compelling argument to advance an alternative approach to better understand and interpret the source for this historical uncertainty. We contrast the phenomenology against theoretical expectations of a massive, diffuse, and spatially extended halo thought to surround the Galaxy. Rather, the case is strengthened for a physically compact, high energy density object having well defined velocity substructure and unique dynamic signatures that present challenges for the current paradigm within and beyond the Galactic disk (and by extension, the dark matter halo).

In this work, we further extend and strengthen the proposed 'scaling model' in which a galaxy's dynamics can be explained through the lens of classical mechanics and thermodynamics (La Fortune 2019). This work leverages a reinterpretation of two popular galactic scaling relations; Mass Discrepancy-Acceleration Relation (MDAR) and the Radial Acceleration Relation (RAR) in a manner that does not require modified gravitation or dark matter halos (La Fortune 2020). We strictly define a small number of simple scaling parameters and apply them directly to baryonic phenomenology, sidestepping problematic "controversies" based on conformance required by ΛCDM-based cosmology (Weinberg 2015).

We begin with simple definitions for dynamic and virial mass and build a physical model for the Galaxy based on them. Recent observations of the stellar halo and dwarf satellite galaxy distribution provide clear differentiation between the scaling model and the two other paradigms. We complete this analysis with a plot of the Galaxy's physical MDAR and RAR based on Gaia DR2 data.

*Generalized Scaling Relations*
Rather than relying on brute-force approach to extricate scaling relations from dark matter halo simulations, the relations are self-evident and obvious from direct observation. We eliminate confusion prevalent in ΛCDM cosmology by narrowly defining dynamic and virial mass – as they represent two different but inextricably linked *physical* quantities. Although virial mass is fairly well defined in ΛCDM halo simulations, the concept of dynamic mass has been missing.

Below, we provide definitions and simple scaling relations with connection of Newtonian law:

- Observed Mass: Dynamic: $M_{Dyn}$ (determined from rotation curves) and virial: $M_{Vir}$ (obtained from stars near or at escape speed via a velocity cut and can include the hypervelocity cohort), and $M_{Bar}$ (either by fixing mass discrepancy or by observation)
- Mass Discrepancy: $D=M_{Obs}/M_{Bar}$ ($M_{Obs}$ can either be *dynamic* or *virial* depending on the specifics). An equivalent form is $D=V_{Obs}^2/V_{Bar}^2$ (using same precautions as mass).
- Galactic acceleration: $g_{obs}=V_{Obs}^2/R$ (again requires knowledge of the mass type linked to observed velocity) and $g_{Obs}=(DM_{Bar})G/R^2$ (subject to the specific definition of $M_{Obs}$ or $V_{Obs}$).

From this perspective, care must be taken to ensure that the data is appropriate for the solutions sought. A source of inaccurate simulation performance stems from employing disk rotation fits with conversion to dark halo mass without a *truly* physical foundation to make such extrapolations. The net result has been a proliferation of a multitude of dark matter halo mass definitions. The issue has become so prevalent that there is a movement underway to better standardize halo parameters (Diemer 2020). For this work, we define two simple and modest 'scaling model' mass terms.

*Scaling Model Definition – Dynamic Mass ($M_{Dyn}$)*
For the scaling model, we strictly define $M_{Dyn}$ as the mass associated with or obtained from galactic rotation curves. This dynamic mass is based on the dimensionless Peebles spin equation $\lambda=J\sqrt{E}/GM^{5/2}$ (Peebles 1971). Employing this equation and nominal disk properties, Peebles demonstrated galactic dynamic mass was five-times greater than the naïve expectation. Today, this is termed the mass discrepancy ratio and still remains an important scaling ratio in the study of galactic dynamics. Although virial theory was used to determine the spin equation, in practical application, it has been associated with disk rotation dynamics and not specifically virial mass per se. The Peebles spin equation is within the realm of classical mechanics and provides a very straightforward definition of dynamic mass as originally intended.

*Scaling Model Definition – Virial Mass ($M_{Vir}$)*
We make several conservative assumptions with regard to disk galaxies. The first is that they are classically-defined thermodynamic objects subject to known physical laws. These objects exist in long-lived stable quasi-equilibrium states within their immediate environs (the classic "system-surroundings" situation). As a central tenet of the scaling approach, each galaxy must have a clearly defined virial mass ($M_{Vir}$) with phenomenology consistent with a *truly* thermalized object. Some attributes include high kinematic stochasticity, high dispersion, and a broad orbital eccentricity distribution. Note that this does not apply to the highly circularized, dynamically constrained orbits dominated by ordered motion. With correct definition, we show galactic mass discrepancies tightly range between "Peebles" dynamic mass (D~5) and virial mass (D~12). There is no ambiguity between the two definitions as they are determined via two entirely different methods.

With advent of ΛCDM cosmology, these simple definitions have become less meaningful with 'dark' terms prevailing today over their truly physical equivalents. As such, there are now convenient correlations between scaling $M_{Dyn}$ (as measured by rotation curves) and associated dark halo 'virial' mass ($M_{200} \approx M_{Vir}$) indicating that a clear distinction still exists between them (Yu 2020). Rather than dark matter halo properties quantifying $M_{Vir}$, the scaling approach employs the classical Maxwell-Boltzmann (M-B) probability distribution (King III 2015) (La Fortune 2017).

We demonstrate that this distribution provides a compelling physical model that just not only describes the virial dynamic, but *explains* much of the phenomenology being discovered in our Milky Way galaxy without resorting to 'new physics.' One example is the Galactic M-B virial signature with orbital characteristics that include stochastic kinematics, isotropic orbits, anisotropy coefficient β≈0, and mean eccentricity *e*≈0.55. The kinematics may be altered by the Galaxy's gravitational potential, angular momentum, and ongoing secular processes (star formation, etc.) but it is anticipated that the presence of the M-B distribution should still remain recognizable and thus quantifiable.

Motivated by this particular velocity probability distribution, we introduce standard definitions to fix the magnitude and width of the profile. We consider the observed peak velocity functionally equivalent to the "most probable" velocity ($V_{MP} \rightarrow V_{Esc}$) at radius $R_P$ where the peak (determined from component velocity dispersions) is observed. More importantly, it also establishes $M_{Vir}$ as a separately calculated mass parameter (scaling $M_{Vir}$) that is decoupled from mass determined from rotation velocities ($M_{Dyn}$). In the next sections, we pressure-test the scaling model against high quality measures of the Galactic stellar halo within 50 kpc and satellite galaxies inhabiting the low acceleration regime well beyond the physical disk radius.

*Milky Way Scaling Model vs. Stellar Halo Observations to 50 kpc*
In this section we compare the scaling model to a study using data obtained from the Slone Digital Sky Survey, prior to Gaia second data release the following year (Williams 2017). This method relies on a "best fit" Spherically-symmetric Power Law model (SPL) using a groomed sample of MSTOs (Main Sequence Turn-Off), BHBs (Blue Horizontal Branch) and K-giant stars to trace the Galactic profile. To mitigate disk contamination, a velocity cut was applied to assure the sample represents the stellar halo with the model extending to 50 kpc.

In Figure 1 below, we reproduce Williams' result (solid gray line and bracketed points) against the scaling template for the Milky Way galaxy. Since results were derived from stellar velocities near or at $V_{Esc}$, the upper point represents $M_{Vir}$. The SPL profile suggests a low total (enclosed) Galactic mass $M_{Tot}=0.29 \times 10^{12} M_\odot$ (gray solid) which is inconsistent with D=12.1 as the effective upper bound in mass discrepancy. From a scaling perspective, it appears that the Williams SPL $M_{Tot}$ value may be related to the Galaxy's dynamic mass $M_{Dyn}$ within 50 kpc.

Below, the generalized scaling template for the Galaxy is illustrated. (Observed) features include a constant disk circular velocity $V_C=230$ kms$^{-1}$ that reaches to the disk edge $R_D=40$ kpc (open red circle). Virial related geometry $R_P=23$ kpc and peak virial velocity is denoted by $V_{MP}=432$ kms$^{-1}$ linked to the M-B velocity probability distribution (red cross). The empirically derived scaling masses are $M_{Dyn}=0.5 \times 10^{12} M_\odot$ and $M_{Vir}=1.0 \times 10^{12} M_\odot$ corresponding to the disk and virial components, respectively.

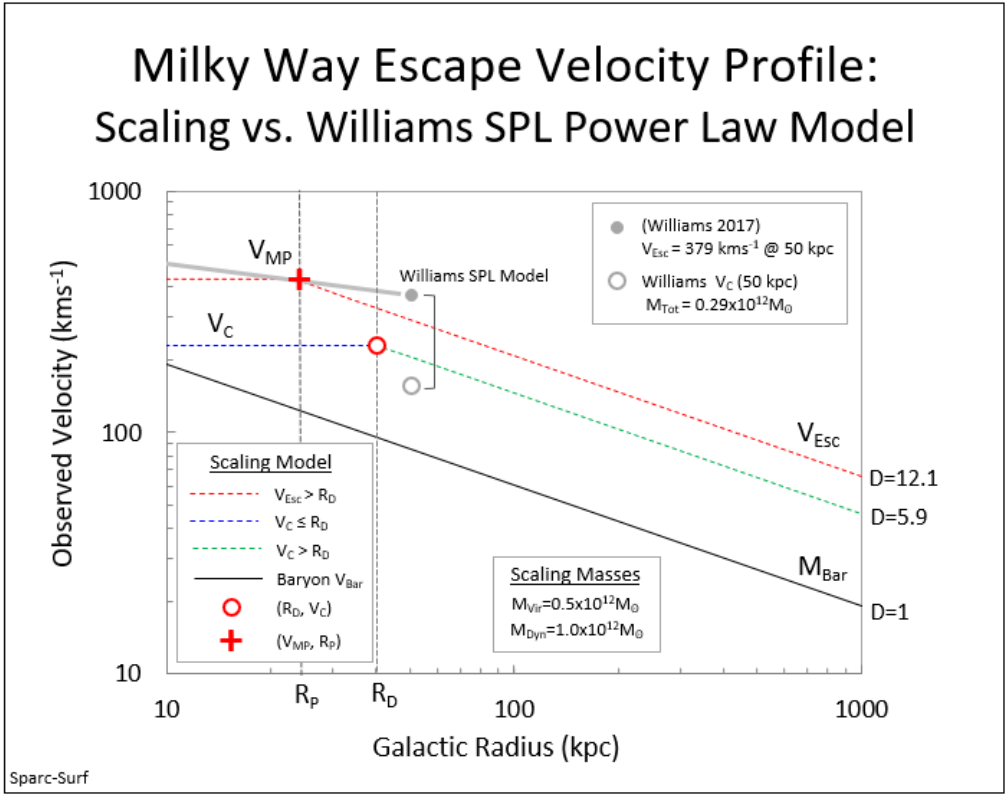

*Figure 1: Generalized Scaling Template for the Milky Way and observed stellar halo velocities as a function of radius – see key for details. The Williams virial mass estimate $M_{Vir}=R_{50}V_{Esc}^2/G$ is $M_{Vir}=1.67 \times 10^{12} M_\odot$ (gray point). Data source - (Williams 2017)*

In the above figure, the scaling and SPL models are in good agreement. The solar/local escape velocity used in the SPL model was fixed at $V_{Esc}(R_\odot)=521$ kms$^{-1}$, precisely the same value obtained from the scaling model. At the scaling virial radius $R_P=23$ kpc, the SPL and scaling escape velocities are $V_{Esc}=428$ kms$^{-1}$ and $V_{MP}=432$ kms$^{-1}$, respectively. Per scaling definition, the SPL $V_{Esc}=379$ kms$^{-1}$ result corresponds to a very high enclosed virial mass $M_{Vir}=1.67 \times 10^{12} M_\odot$. In order to maintain mass discrepancy below D=12.1, $M_{Tot}$ would need to be fifty-percent baryonic, an unreasonably high value. While not known at the time of this study, the Williams data set traverses two separate, distinct dynamical regions (inside and outside $R_P$) and is not accounted for using the single power law assumption.

Turning to scaling parameters, the nominal Galactic circular velocity is $V_C=230$ kms$^{-1}$ (blue dash) for the disk truncating at $R_D$ (open red circle) which is considered the disk's dynamical edge. Beyond $R_D$, a conventional Keplerian decline follows (green dash) per the Newtonian prescription. The virial-sourced Keplerian virial escape velocity profile (red dash >$R_P$) parallels the $V_C$ curve offset by $\sqrt{2}V_C$. Physically, this offset enables escape velocities to align, resulting in a singular global dynamic beyond $R_D$. Along the entire Keplerian trace, this alignment keeps both curves within 10 kms$^{-1}$ of each other. To distant observers, the Galaxy would appear as a point mass object with total mass $M_{Vir}=1.0 \times 10^{12} M_\odot$, not the combined sum of $M_{Dyn}$ and $M_{Vir}$.

A more recent study using Gaia DR2 stellar data and a method similar Williams provides comparable results and includes an excellent discussion on this topic (Koppelman 2020). In the next section, we investigate kinematics measured from the recent Gaia DR2 survey of the Milky Way's dwarf satellite galaxy population.

*Milky Way Scaling Model and Gaia DR2 Dwarf Galaxy Satellite Observations to 300 kpc*
The Milky Way is surrounded by a small population of satellite galaxies that trace the global potential to beyond 300 kpc. Precision measures are now available to determine if proposed scaling constraints are appropriately motivated and accurately portrays Galactic structure and kinematics.

Figure 2 presents the orbital characteristics for a sample of the Galaxy's dwarf galaxy satellites obtained from Gaia DR2 (Fritz 2018). We include his best fit NFW dark matter halo virial mass $V_{Esc}$ profile (red solid) and the Williams SPL modeling results (gray points) for perspective. Fritz separated satellite data into two categories; a cohort having high quality orbital parameters (blue solid) and a smaller cohort with larger velocity errors (open blue circles). Focusing on high quality satellite galaxy data, visual inspection of Figure 2 reveals the majority are constrained to a narrow region between $M_{Dyn}$ D=5.9 (green dash) and $M_{Vir}$ D=12.1 (red dash) from $R_D$ outward to 300 kpc.

Note that inside $R_P$ the $V_{MP}$ and $V_{RMS}$ value are constant (horizontal) indicative of a pseudo-isothermal regime, while at $R_P$ and beyond satellites decline in Keplerian fashion. The gray shaded region along and outside $V_{MP}$ >$R_D$ represents the effective $V_{Esc}$ profile for the Galaxy. We include the virial $V_{RMS}$ value to express the relatively wide range of velocities possible within this particular M-B distribution. Without understanding the motivation and nature of the physical dynamic, accurate measures of galactic $M_{Vir}$ and $V_{Esc}$ are difficult to attain. This can lead to the belief that persistent 'uncertainty' in these estimates are a function of the measurements and/or models when in reality, it is a reflection of the stochasticity inherient in the M-B distribution itself.

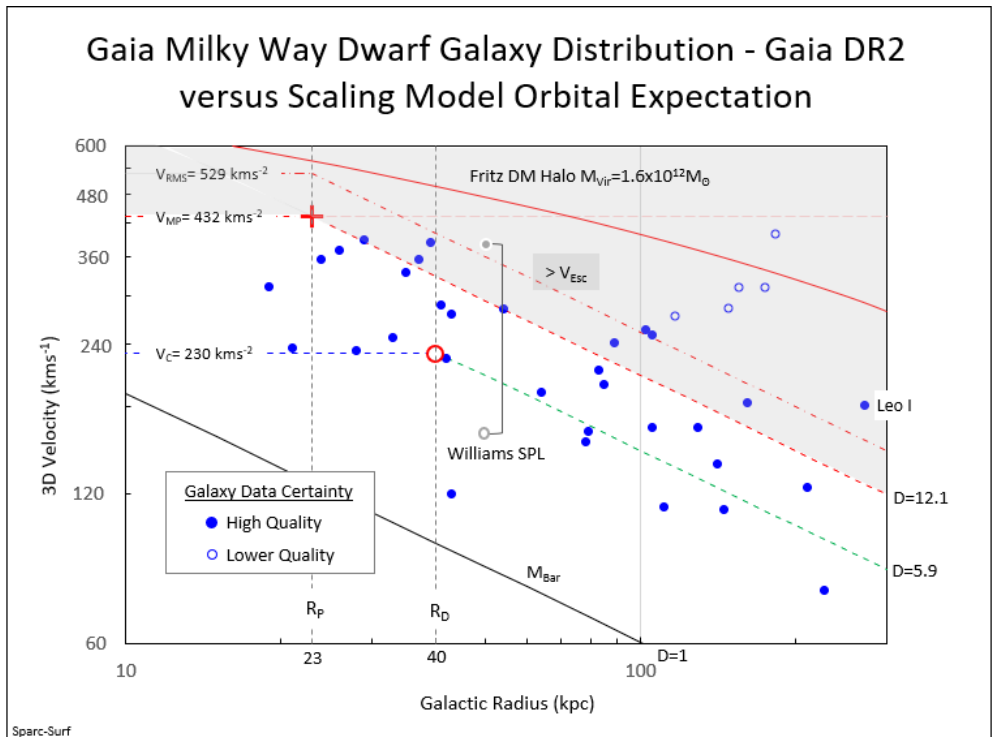

*Figure 2: Scaling Model and Observed Velocities as a function of radius for the Milky Way Galaxy. SSDS Stellar halo estimates (gray) and Gaia DR2 dwarf satellite galaxies (solid and open blue points). Note narrow band of satellites between D=5.9 and 12.1 declining in Keplerian fashion (green and red dash > $R_D$). $V_{MP}$ fixes the geometry and magnitude of the effective escape velocity profile. Fritz estimated dark matter halo virial mass escape velocity curve (red solid). Leo I has been identified for later discussion. Data source – (Fritz 2018)*

The data in Figure 2 reveal an *inconsistency between the satellite distribution and expectation for a featureless diffuse dark matter halo*. Considering that halos are thought to be completely virialized to roughly 250 kpc (near the right edge of the plot) the satellite distribution should be more uniformly distributed in the halo to $V_{Esc}$ (red solid). We find a *robustly constrained R-V envelope following a 1/√r profile within the dark matter halo.*

The above figure shows a strong dynamic 'partition' at $R_P$ evidenced by the very sharp reduction in satellite galaxy counts interior to $R_P$ (as opposed to the large number of satellites populating the disk inside $R_D$). This observation is in conflict with halo dynamics with satellite numbers smoothly declining with decreasing radius due to dark matter interaction between the host halo and satellite galaxies (aka, luminous subhalos).

We also note that the disconnect is even more stark with predicted satellite trajectories showing orbital pericenters almost entirely absent within virial radius $R_P$. The result infers that this dynamic partition may enclose a "forbidden zone" that effectively destabilizes orbital trajectories and sweeps satellites out over time.

In Figure 3 below, we provide further substantiation of Galactic substructure from a more recent analysis from the Gaia DR2 sample (Fritz 2020). For this view, satellite velocities are plotted in linear fashion to better visualize the distribution, constraining dynamic, and Keplerian declines. In this figure, a tight correlation is obtained between the 'High Quality' sample (blue solid) and local escape velocity (red dash) per scaling expectations. This plot shows that $V_{MP}$ can be considered a reasonable parameter to estimate $M_{Vir}$. We include revised mass estimates from Fritz (gray and white filled dark gray circles) that better 'fit' the data. With revised computation, his mass estimates now agree with the scaling method and results.

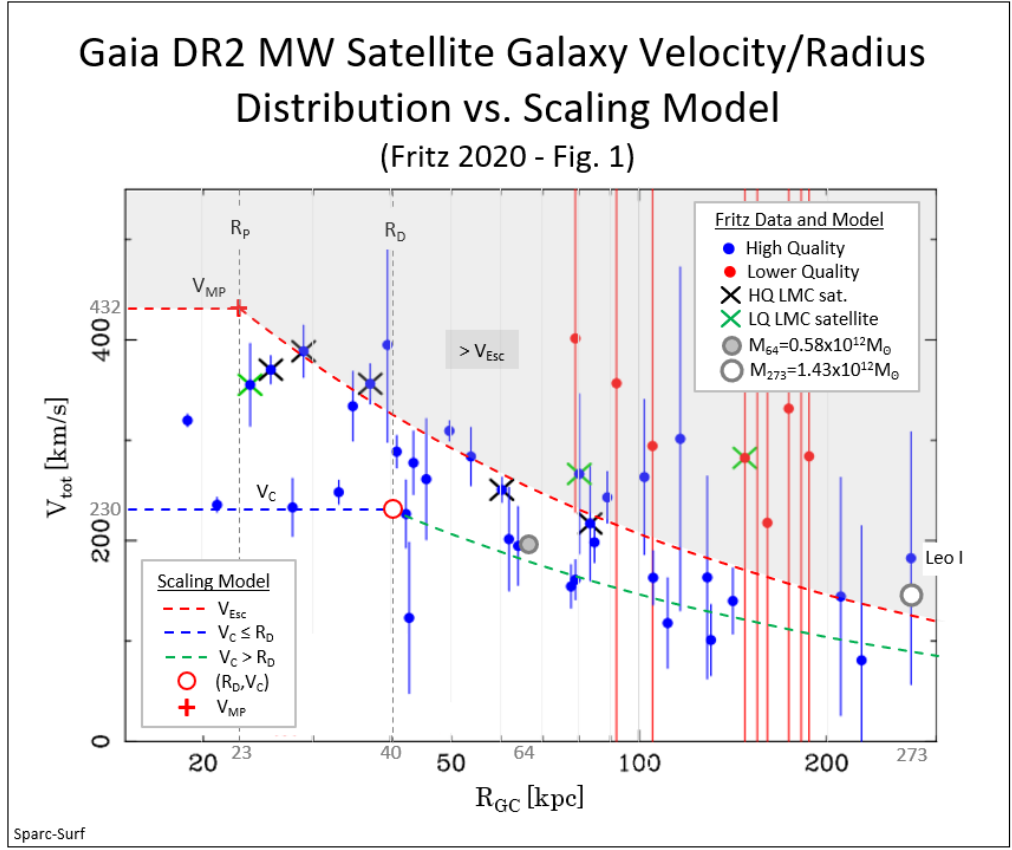

*Figure 3: Observed Milky Way satellite data from Gaia DR2 – total velocity as a function of galactic radius with scaling parameters superimposed. The majority of the 'high-quality'' satellite sample (blue data with error bars) do not significantly exceed the global escape velocity from 18 to 300 kpc (red dash). Fritz's 2020 revised halo mass estimates (filled and open gray circles) are in better agreement with the scaling model, but is still overestimated based on Gaia DR2 satellite galaxy kinematics. The Fritz $M_{64}$ mass is twice that obtained by Williams SPL stellar halo model. Source – (Fritz 2020) Fig.1- permission required.*

In Figure 3, the estimated virial mass of the Fritz dark matter halo is $M_{Vir}=1.51 \times 10^{12} M_\odot$ consistent with the M-B $V_{RMS}$ parameter, but still significantly overestimated. Fritz's unusually low $M_{64}$ mass value may be an artifact of modeling the Keplerian decline with a dark matter halo model or blend of disk dynamics and virial support.

There are two sources of mass estimate uncertainty inherient in ΛCDM halo models. The first is the intrinsically wide range of the masses possible due to the cosmological assumption >99% of all dark matter subhalos (satellite galaxies) must be bound within the virialized host dark matter halo. The Leo I satellite highlighted in the figure as it is often leveraged to establish dark matter halo mass. If removed from halo model fits, it can be shown that the halo would provide a more reasonable fit in the outer perimeter.

The scaling model interpretation places Leo I squarely within the M-B distribution and does not overly influence the physical $M_{Vir}$ estimate as obtained from the entire sample. The fact that dark matter halo mass rests on the kinematics of an 'influential' individual satellite as dictated by theory may need to be revisited.

As an example of this uncertainty, a recent compilation of Galactic dark halo 'virial' mass estimates gives best-fit estimates spanning the wide interval [0.5, 2.0] x$10^{12}$M$_\odot$ (Wang 2020). In addition to unacceptable uncertainty between estimates, normalization (magnitude) is also amiss. This was demonstrated in two comprehensive studies comparing dark matter/baryon ratios obtained from state-of-the-art hydrodynamical simulations and rotation curve fits for a subset of massive spiral galaxies culled from the SPARC data set (Posti 2019) (Marasco 2020). Both studies showed these simulations yielding disproportionately high disk dark matter fractions two-to-four times what is required to match the phenomenology.

Not surprising was that SPARC massive spiral galaxies contain their full complement of baryons (D=5.9) which did not allow for any significant dark matter content in the inner disk region. Per the scaling prescription, nearly all galaxies contain their full complement of baryons <0.125$R_D$ (where D≈1). While this scaling expectation directly contradicts simulation results, it is in agreement with modified Newtonian gravitation theories that only have baryons as their motivating component (Milgrom 1983). This significant mass discrepancy (especially within the inner disk region) is perhaps the most pervasive small-scale 'controversy' facing ΛCDM cosmology today. The oft cited dark matter halo 'cusp-core' dilemma is just a particular symptom of a chronic problem that is now being repaired using ad hoc baryon physical processes to move dark matter into its 'correct' location.

Observational confirmation for true physical virialization of the Milky Way >$R_P$ has been given by Fritz with the estimated anisotropy coefficient being β=-0.21 (+0.37 -0.51) consistent with an isotropic distribution (β=0) and a mean eccentricity $e$≈0.57. Inside $R_P$, the dynamic is much different with an investigation consisting of Gaia DR2 RR Lyrae stars between 5 and 25 kpc demonstrating the majority of this stellar cohort orbiting radially with an average anisotropy coefficient β≈0.9 (Iorio 2020).

The dichotomy in dynamics interior and exterior to $R_P$=23 kpc can be physically explained on a simple premise: the high anisotropy observed inside $R_P$ is an artifact of the spherically-shaped virial mass 'shell.' Interior to $R_P$, the kinematics are complex due to the high constant gravitational potential and little radial field due to the massive thin shell nature attributed to the virial surface. This unique dynamic is rather benign, permitting 'rectilinear-like' motion and long-lived bulk/group transport to exist. Moreover, the virial shell acts as a "trapping" mechanism. For example, orbits passing through $R_P$ in outward trajectories will be pulled back toward the virial surface on pseudo-ballistically while those orbits penetrating inward experience an abrupt reduction in the gravitational field. These trajectories will tend to amplify radial anisotropy with each pass (unless perturbed or ejected from the Galaxy due to severe destabilization). It is this unique dynamic that is responsible for highly radial orbits and their longevity as evidenced by the existence of the Gaia Sausage stellar distribution and streams found to be confined within the virial dynamic.

We make a special point to discuss the lower quality satellite galaxy sample (red points with extended error bars) identified by Fritz and shown in Figure 3. Rather than dark matter halo 'splashback' mechanisms offered by Deason, we offer two *physical* alternatives (Deason 2020). We find that either these satellites have been subjected to intense orbital destabilization from passing into and out of the virial partition or they are in the process of accreting close to terminal entry speed $V_{MP}$. Deason places the true edge of the Galaxy as determined by simulated halo results at 292±61 kpc versus $R_D$=40 kpc based on a modest interpretation of the latest precision astrometrics. It should be noted that there is no evidence for a 'splashback' radius or edge that has been measured – these radial signposts only exist in computer code (with 'low-quality' satellites used in the same fashion as Leo I – overleveraged).

*Milky Way Mass MDAR and RAR Analysis – Gaia DR2 Satellite Galaxy Data*

No analysis of the Galaxy would be complete without an MDAR plot as demonstrated by McGaugh, Lelli, and Schombert – termed 'MLS' in figures and text below (McGaugh 2016). The purpose of this diagram is to provide a convenient method to describe the RAR based on baryon and dynamic acceleration ratios in their version of modified Newtonian gravity. We provide this analysis shown in Figure 4 because it communicates an important aspect of galactic dynamics in a 'scale free' format, and thus is perfectly suited for our model based on phenomenal scaling relations. We include the universal RAR (gray solid) extending to *and* beyond the edge of the plot). The RAR is expressed in the following equation as fit by the SPARC data set:

$$MLS\ g_{Obs} = \frac{g_{bar}}{1 - e^{-\sqrt{g_{bar}/g_†}}} \quad with \quad g_† = Milgrom\ Constant\ "a_0"$$

Unlike most MDAR figures, we provide mass discrepancy iso-contours for D=1 ($M_{Bar}$), D=5.9 ($M_{Dyn}$) and D=12.1 ($M_{Vir}$) to highlight important aspects of the aforementioned acceleration constraints.

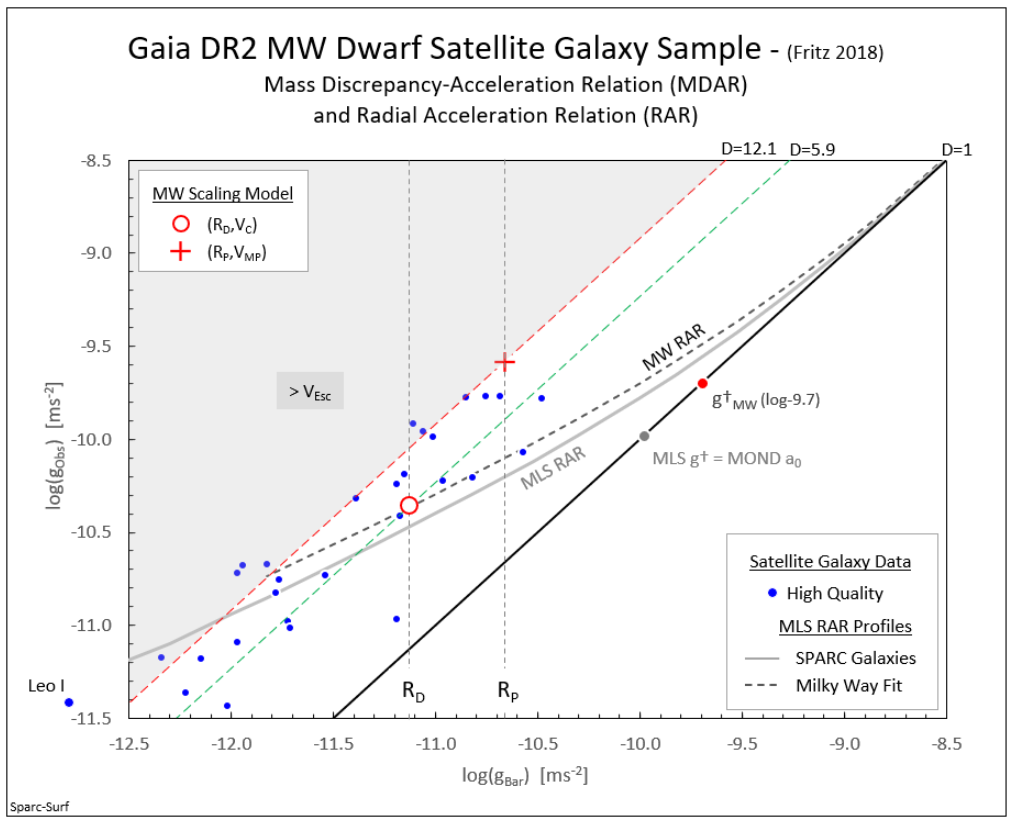

*Figure 4: Observed Mass Discrepancy-Acceleration Relation of Gaia DR2 obtained dwarf galaxy satellites (blue solid). Included are the MLS RAR profile (solid gray) and observed Galactic fit (black dash). Satellite galaxy Leo I is highlighted as it is often used to fix dark matter halo mass equivalent to $M_{Vir}=1.6 \times 10^{12} M_\odot$ ($D_{Vir} \sim 20$). Sources – (Fritz 2018) (McGaugh 2016)*

In Figure 4, we find the MLS acceleration constant g† on D=1 (gray point - identical to Milgrom's constant $a_0$) differs from the one calculated for the Galaxy (red point on D=1). Another issue with the RAR as touted by MLS is that the satellite sample does not follow nor extend beyond D=12.1 as expected in their theory of modified gravitation being a natural law.

Currently, there is increasing interest in comparing the MLS RAR against observed accelerations beyond individual galaxies. Two studies have been published for galaxy clusters and groups with the aim to determine if the quoted acceleration constant ($g\dagger \equiv 1.2 \times 10^{-10}$ ms$^{-2}$) is truly universal irregardless of structure or internal dynamics (Chan 2020) (Tian 2020).

*Summary*
We have examined the physical geometry and kinematics of the outer region of the Milky Way galaxy and demonstrate that the scaling model offers a new approach to describe and explain newly revealed substructure obtained from precision astrometrics. The scaling model is physically motivated, offering a new perspective that does not rely on dark matter or modified gravity to explain these recently revealed dynamic features. There is every expectation that subsequent quality observations will further strengthen the model, while those models that rely on 'best fit' approaches (modeling by exception) will continue to struggle with internal inconsistencies that run counter to our best available observations.

*Acknowledgements*
We thank those assisting in the development of this proposal and permissions to reproduce previous works. We thank arXiv for its preprint platform and service.